%% file: main.tex
\documentclass[12pt,twoside,a4paper]{article}
\usepackage[hmargin=25mm,vmargin=20mm]{geometry}
\usepackage{hyperref}

\usepackage{epsfig,graphicx,parskip,setspace,tabularx,xspace,todonotes}

\graphicspath{{images/}}

\def\first{{\it 1})\xspace}
\def\second{{\it 2})\xspace}
\def\third{{\it 3})\xspace}

\newcommand{\ie}{i.e.\@\xspace}

\newcommand{\etc}{etc.\@\xspace}

\usepackage{fancyref}	
\newcommand*{\fancyreflstlabelprefix}{lst}
\fancyrefaddcaptions{english}{%
  \providecommand*{\freflstname}{listing}%
  \providecommand*{\Freflstname}{Listing}%
}

\frefformat{plain}{\fancyreflstlabelprefix}{\freflstname\fancyrefdefaultspacing#1}
\Frefformat{plain}{\fancyreflstlabelprefix}{\Freflstname\fancyrefdefaultspacing#1}

\frefformat{vario}{\fancyreflstlabelprefix}{%
  \freflstname\fancyrefdefaultspacing#1#3%
}
\Frefformat{vario}{\fancyreflstlabelprefix}{%
  \Freflstname\fancyrefdefaultspacing#1#3%
}
\vrefwarning

\usepackage{amsmath}
\usepackage{caption}
\usepackage{subcaption}
\usepackage{booktabs}
\usepackage{multicol}
\usepackage{multirow}

\usepackage{float}
\usepackage[figuresright]{rotating}
\usepackage{tabularx}

\begin{document}
\setcounter{page}{1}

\title{Hearing your touch:\\ A new acoustic side channel on smartphones}
\author{Ilia Shumailov \and Laurent Simon \and Jeff Yan \and Ross Anderson}
\date{} 

\maketitle 

\begin{abstract}
{We present the first acoustic side-channel attack that recovers what users type on the virtual keyboard of their touch-screen smartphone or tablet. When a user taps the screen with a finger, the tap generates a sound wave that propagates on the screen surface and in the air. We found the device's microphone(s) can recover this wave and ``hear'' the finger's touch, and the wave's distortions are characteristic of the tap's location on the screen. Hence, by recording audio through the built-in microphone(s), a malicious app can infer text as the user enters it on their device. We evaluate the effectiveness of the attack with 45 participants in a real-world environment on an Android tablet and an Android smartphone. For the tablet, we recover 61\% of 200 4-digit PIN-codes within 20 attempts, even if the model is not trained with the victim's data. For the smartphone, we recover 9 words of size 7--13 letters with 50 attempts in a common side-channel attack benchmark. Our results suggest that it not always sufficient to rely on isolation mechanisms such as TrustZone to protect user input. We propose and discuss hardware, operating-system and application-level mechanisms to  block this attack more effectively. Mobile devices may need a richer capability model, a more user-friendly notification system for sensor usage and a more thorough evaluation of the information leaked by the underlying hardware.}
\end{abstract}

\input{sections/intro}
\input{sections/background}

\input{sections/evaluation}
\input{sections/countermeasures}

\input{sections/limitations}
\input{sections/related_work}

\input{sections/conclusion}

\bibliographystyle{plain}
\bibliography{catmax.bib}

\end{document}

%% file: sections/intro.tex
\section{Introduction}

An extensive body of research has shown how to exploit built-in smartphone sensors 
to extracts users' personal information through side channel attacks; see~\cite{ALHAIQI2013989,Andriotis:2013:PSS:2462096.2462098, aviv2012side,Aviv:2012:PAS:2420950.2420957, cai2011touchlogger, Goller2015, Li:2016:CMP:2976749.2978397,  Negulescu:2015:GCI:2702123.2702185,   7368569, Simon:2013:PSI:2516760.2516770, DBLP:journals/popets/SimonXA16,  Xu:2012:TIU:2185448.2185465,   Yan:2015:SPS:2875913.2875934} and~\cite{DBLP:journals/corr/abs-1808-10250}. 

A comprehensive survey of such attacks~\cite{Cai:2009:DAS:1592606.1592614} pointed out that side-channel attacks using microphones are less well studied. Acoustic attacks against \textit{physical} keyboards have received some attention; see~\cite{asonov2004keyboard, keyboardacous,  Liu:2015:SKM:2789168.2790122, atmkeypad2009, Zhu:2014:CAU:2660267.2660296, zhuang2009keyboard} and~\cite{DBLP:journals/corr/CompagnoCLT16}. Each key has  different physical characteristics or defects which enable classification. Acoustic attacks on \textit{touchscreen soft} keyboard are inherently more difficult, since each finger tap happens on the same surface. Some active attacks were explored by~\cite{DBLP:journals/corr/abs-1808-10250}; the previous work on passive acoustic attack on virtual keyboards is by~\cite{Simon:2013:PSI:2516760.2516770}, who used the audio for tap detection, not for classification. 

In this paper, we present the first fully passive acoustic side channel attack against \textit{touchscreen virtual keyboards}. 
When a user enters text on the device's touchscreen, the taps generate a sound wave. The device's microphones can recover the tap  and correlate it with the keystroke entered by a victim. The spying app may have been installed by the victim herself, or by someone else, or perhaps the attacker gave the device to the victim with the app pre-installed -- there are several companies offering such services, such as mSpy~\cite{Brewster17}.
We also assume the app has microphone access. Many apps ask for this permission and most of us blindly accept the list of demanded permissions anyway~\cite{Felt:2012:APU:2335356.2335360}.

The only previous work we could find on passive acoustic attacks against virtual keyboards is by Narain et al.~\cite{narain2014single}, who claimed that stereo microphones are more effective than other sensors. From their evaluation, however, it is hard to infer the accuracy of their attack, and they assumed that the keyboard reports when a tap happens, which is not in general the case. 

We improved on this early work by overcoming a number of challenges, including extraction of relevant features, reliable detection of finger taps in noisy environments, and re-synchronization of raw signal data (\fref{sec:tapdetection}).
To make the attack stealthier and more practical, our spying app is trained offline; so there is no need to train the model with the victim's data. 
The only constraint is that the attacker to has train the model with the same smartphone or tablet. The whole attack pipeline is presented on \fref{fig:attack_pipeline}.


Through a study conducted with 45 participants, we show that an attacker can successfully perform PIN inference attacks (\fref{sec:evaluation}). We recover 31 out of 50 4-digit PINs in 20 attempts. We also show that keyboard inference attacks are feasible. 
For example, out of 27 passwords, we recover 7 words on the phone and 19 on the tablet within 10 attempts. 
This illustrates the hazards of reasoning about smartphone sandboxing given the complexity of modern platforms, as well as the need for a more realistic threat model for modern hardware. 

We discuss possible countermeasures in \fref{sec:cm}. We propose a number of steps that app authors can take to protect against such attacks, even in the absence of better operating-system support. We then discuss more effective measures that could be built into the operating system or hardware.

We contribute the following:
\begin{itemize}
    \item The first purely acoustic side channel attack against a smartphone \textit{virtual keyboard};
    \item Proof that one can perform accurate time-difference-of-arrival calculations between two microphones on the same phone, revealing where along the axes between two microphones the tap is happening;
    \item An extensive study of PIN and word inference with 45 people on tablets and smartphones;
    \item A comparison of our attack with previous work.
\end{itemize}



%% file: sections/background.tex
\section{Background}\label{sec:background}


\subsection{Tap Detection}\label{sec:tapdetection}

When a user taps the screen of a smartphone, sound waves start propagating not only through the air but also through the device itself. The screen behaves as a plate that is fixed on the sides to the rest of the smartphone body. Once the screen starts vibrating the vibration patterns can be picked up, and when observed with multiple microphones the patterns can in theory be distinguished, as they are unique (up the symmetry in case of two microphones). 

In practice, however, the sampling rate of modern smartphones imposes limits. For example the Nexus 5's screen is made of Gorilla Glass 3~\cite{corning_2015} in which the speed of sound is estimated at $4154.44$ m/s; but the built-in microphones can only sample every $0.0226$ ms, so they sample the energy propagating through the glass every $0.094$ meters.

\begin{figure}
\centering
 \includegraphics[width=0.7\linewidth]{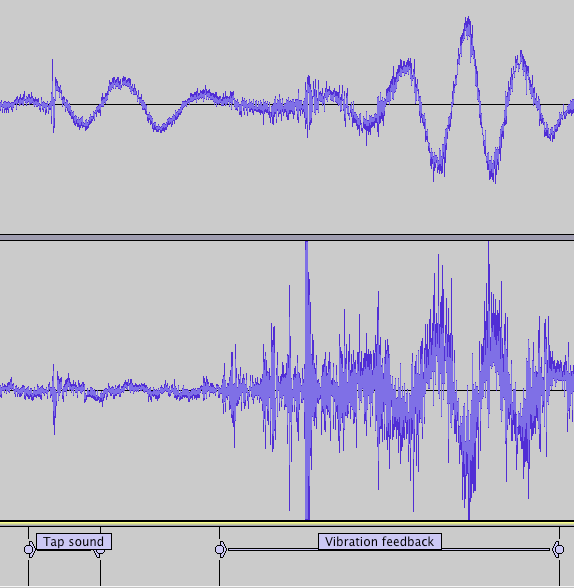}
 \caption{Press on character `f' with a vibration feedback.} \label{fig:vibr_ex}
\end{figure}

The acoustic data captured from a microphone is noisy. The original paper by Asonov and Agrawal that developed attacks against physical keyboards used FFT energy-based thresholding to deal with this~\cite{asonov2004keyboard}, and a similar approach was adopted by many of the follow-up papers~\cite{Berger:2006:DAU:1180405.1180436,zhuang2009keyboard,Zhu:2014:CAU:2660267.2660296}.
This worked quite well in a lab environment, especially since the keyboards they attacked were mechanical and made a lot of noise. However, the real world has a varying signal-to-noise ratio and energy thresholding simply causes a lot of false positives. 
Narain et al. used gyroscope data to detect key taps~\cite{narain2014single}; this is much better then energy thresholding, but still works rather poorly to detect the actual beginning of the tap, given the delays introduced by the OS. We have observed that the timestamps returned by the Android \texttt{elapsedRealtimeNanos} API on a Nexus 5 running Android 5.1.1 are anywhere around $8,000$-$15,000$ samples away from the moment the tap actually happened. In a real-world attack on a target who enters the same PIN a lot of times with two hands, the time difference between taps is swamped by the OS latency.
Simon and Anderson understood this, and used vibration feedback to locate the start of the tap~\cite{Simon:2013:PSI:2516760.2516770}; but vibration can be suppressed by a banking app. 

We therefore set out to solve the problem of purely acoustic tap detection, and observed that the taps have a very distinctive pattern of propagation through the screen and air -- one that is similar for different participants. In particular, we found that high frequencies are observed first for a few samples, followed by a long period of low frequencies.  We still had to deal overlapping signals when the subject typed quickly, but found that both sound and vibration feedback are very distinctive, so it is possible to disambiguate them by subtracting the feedback when the cross-correlation is the highest with the model of the feedback sound. 

\subsection{Time Difference Of Arrival}

\begin{figure}
\centering
 \includegraphics[width=0.4\linewidth]{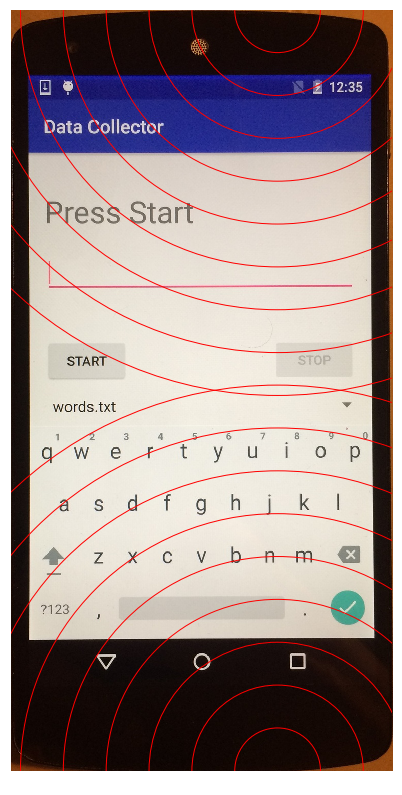}
 \caption{The location of the two Nexus 5 microphones, where the length between the subsequent radiuses is the distance traveled by a soundwave through air during one sampling period.} 
 \label{fig:phone_screen_loc}
\end{figure}

\begin{figure}
\centering
 \includegraphics[width=0.3\linewidth]{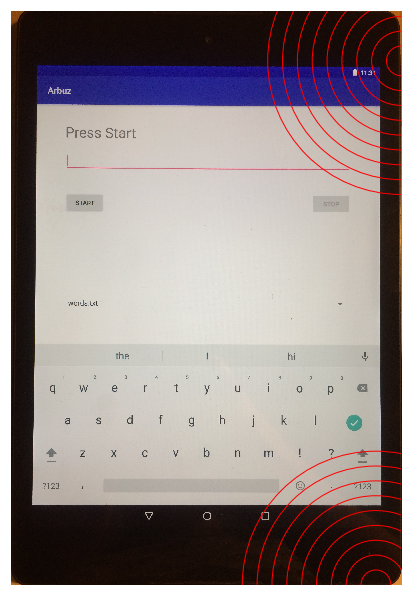}
 \caption{The location of the two Nexus 9 microphones, where the length between the subsequent radii is the distance travelled by a sound wave through air during one sampling period.} 
 \label{fig:tablet_screen_loc}
\end{figure}

\begin{figure}
\centering
 \includegraphics[width=0.3\linewidth]{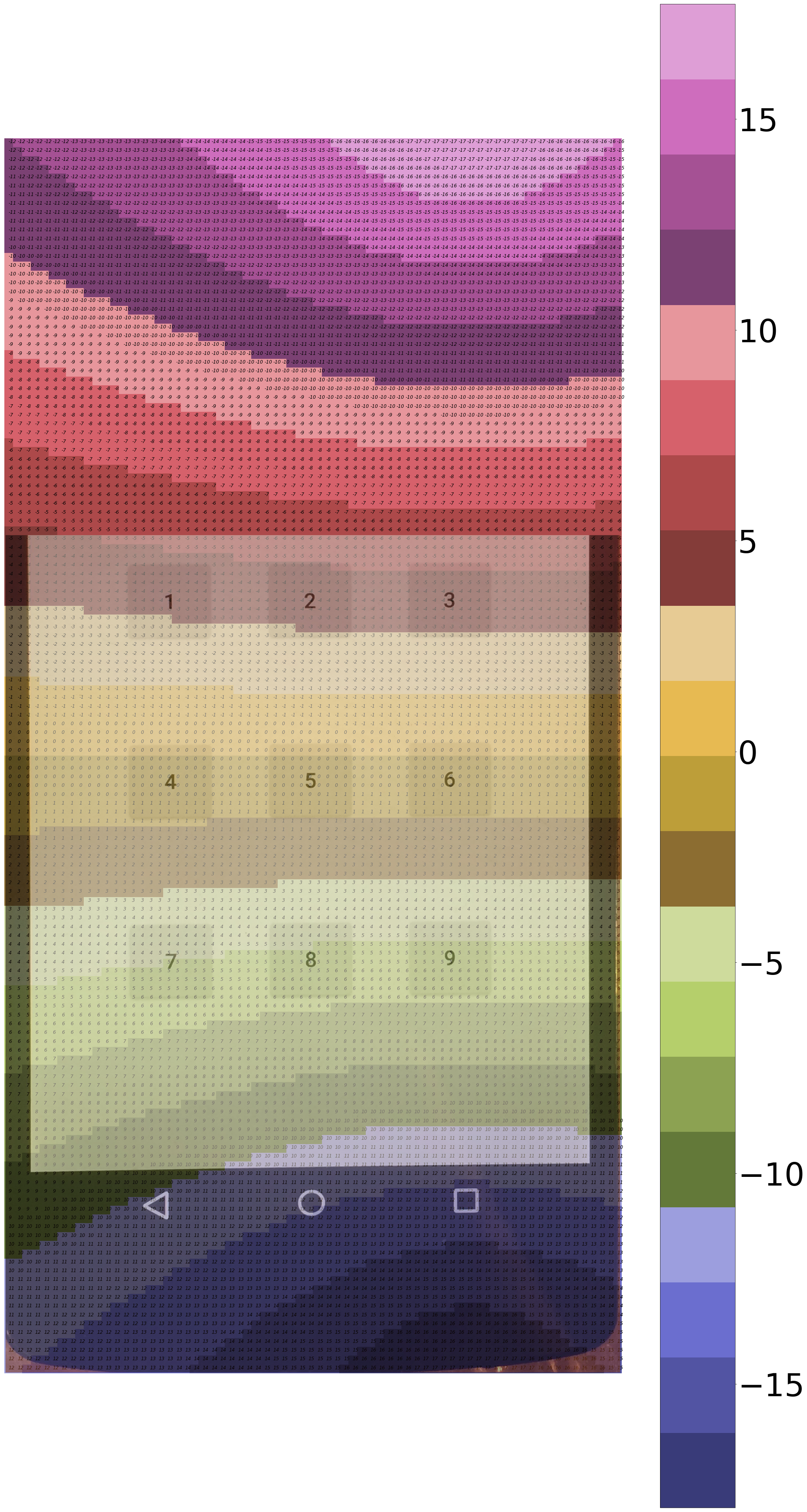}
 \caption{The location of the two Nexus 5 microphones, where the length between the radii around each microphone is the distance travelled by a soundwave during one sampling period.} \label{fig:phone_screen_exp}
\end{figure}

When we have multiple microphones, we can calculate the time difference between the reception of the signal by the bottom one and the top one. 
The theoretical accuracy of time difference of arrival for a Nexus 5 smartphone sampling audio at $44.1$ kHz with two microphones is shown in \fref{fig:phone_screen_exp} -- 
the locations for the Nexus 5 are presented in \fref{fig:phone_screen_loc}, and for the Nexus 9 in \fref{fig:tablet_screen_loc}. This corresponds to samples taken every $8$ mm on the screen.  
The colours correspond to areas where the delay between signal reception by the bottom and the top microphone is constant.
Theoretically, we should be able to distinguish where along the Y-axis the tap is happening when operating in portrait orientation and along the X-axis in landscape orientation. 

\begin{figure*}
    \centering
    \includegraphics[width=0.85\linewidth]{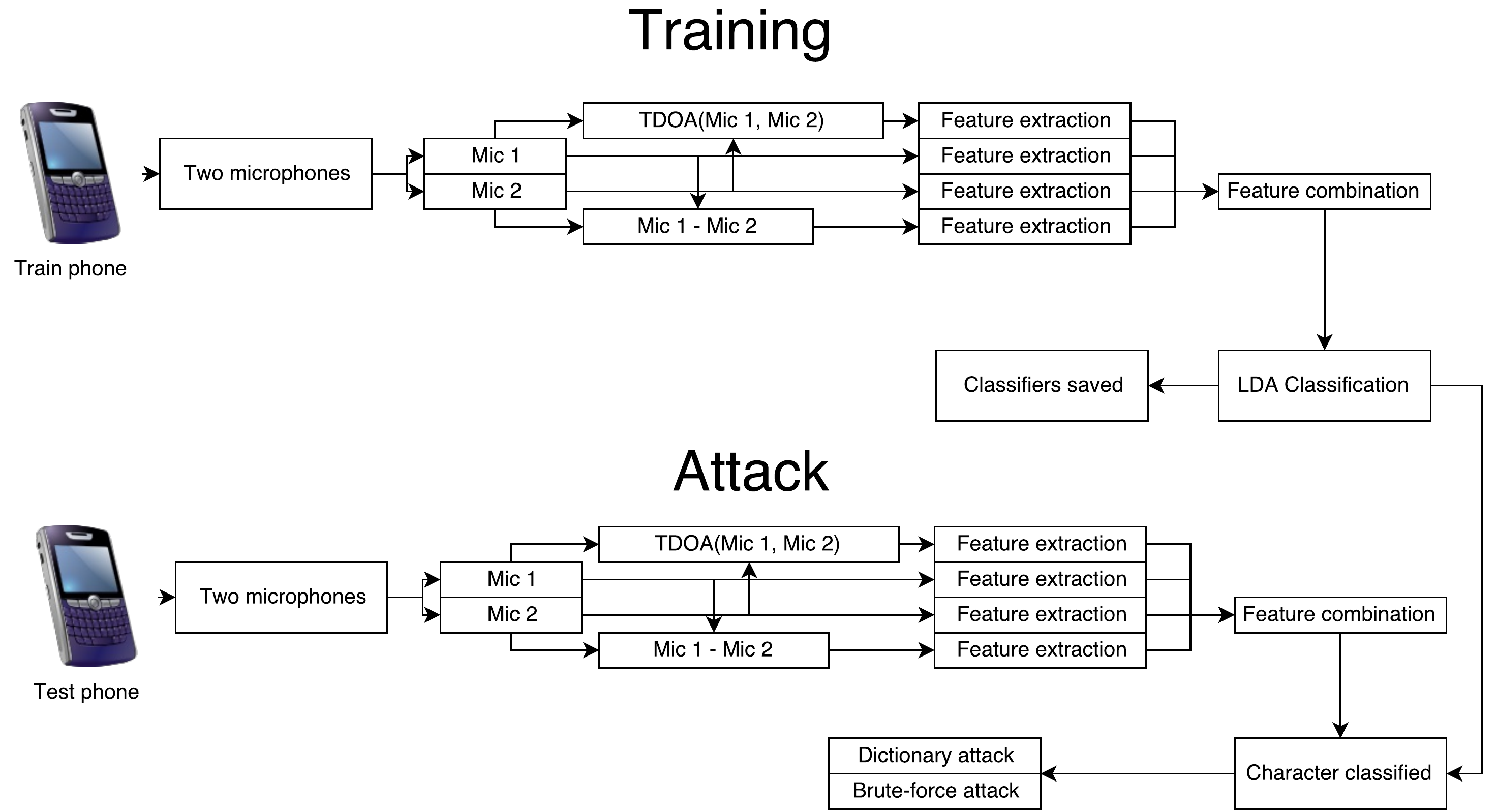} 
    \caption{Attack pipeline.} 
    \label{fig:attack_pipeline}
\end{figure*} 

There are many known methods to compute the time difference of arrival, such as CC, GCC-PHAT, GCC-SCOT, GCC-Roth, GCC-ML, GCC-HBII~\cite{4646820, 1162830}, 
ASDF~\cite{193195, 1162598}, LMS~\cite{1163608,913054}, 
AED~\cite{benesty2000adaptive, 759826, Solo:1994:ASP:184245}, thresholding~\cite{7774858}, ATDC~\cite{103056} 
and combination techniques~\cite{marinescu2013applying}. Each one of these was designed to process particular types of signal with a specific signal-to-noise ratio (SNR). We implemented all of them to see which works best. 

We found that none of them worked reliably on their own. There are several reasons for this. First, the microphone hardware is not ideal and there is some sampling rate skew, so the samples can have a delay between them. 
Second, the environment is noisy and its noise can be difficult to model. Third, we observed that participants often blocked a microphone with their hands when typing. 

After experimenting with several participants' data, we found that preprocessing the data greatly improves the accuracy -- we could use cross-correlation with a signal that is passed through a band-pass filter to achieve reliable results. 
We found that for each tap there is an energy peak at the frequency range of 1,300Hz -- 1,700Hz for both our test devices. So, we decided to use a first-order bandpass Butterworth filter on the signals from both microphones before calculating the time delay. As the time difference is characteristic of the screen location of a user tap, we use it as a feature to input to a classifier. But we also use additional features, as explained next.

\subsection{Acoustics of a tap}

\begin{figure}
\centering
 \includegraphics[width=0.5\linewidth]{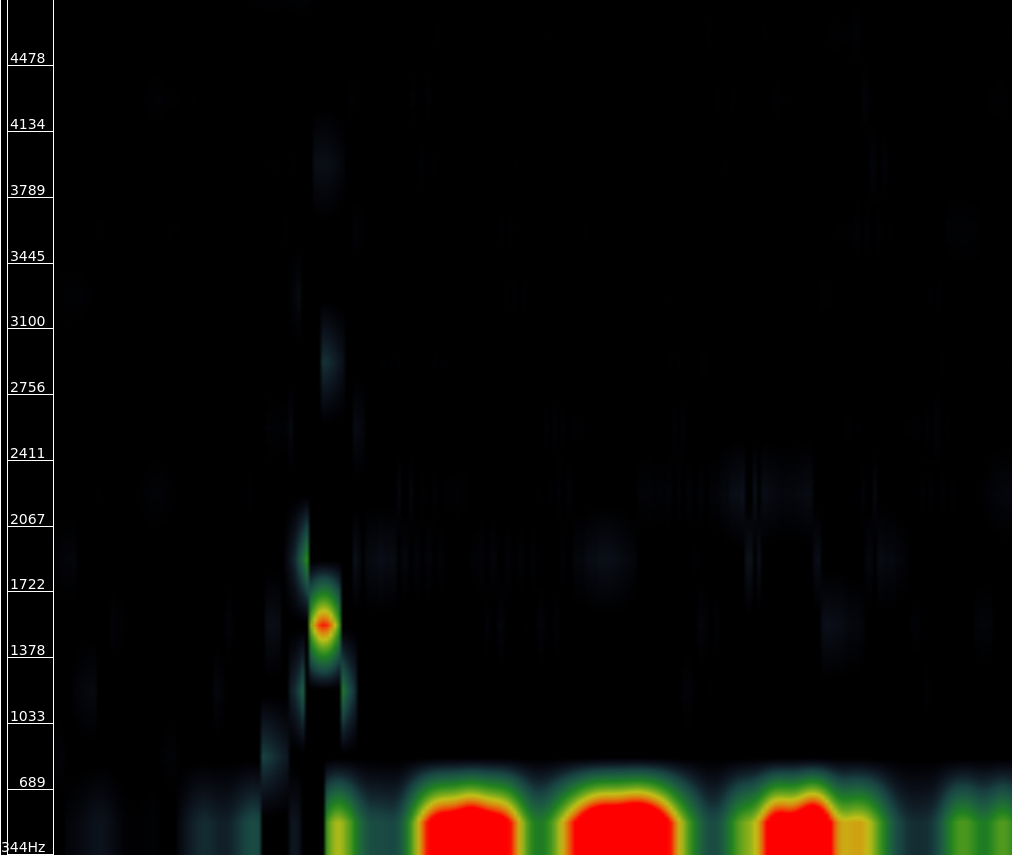}
 \caption{The short-time Fourier transform breakdown of an averaged tap sound.} 
 \label{fig:avg_tap_snd}
\end{figure}

The tap itself is rather short, with an initial burst of about 60 samples at a frequency of about 8000--8500Hz, followed by 4000--4400 Hz, followed by a 60--70 Hz wave for about 1,500 samples (\fref{fig:avg_tap_snd}). Having explored different combinations of features and data-set sizes we found that most of the tap information is actually contained in the first $128$ samples -- in contrast to physical keyboards, where most of the useful information was contained in the release movement of the button.

But with that many samples, we could not efficiently use conventional spectral audio features, such as spectral contrast~\cite{1035731}. We have attempted to over-sample the signal to get sub-sample values for the signal in the given bands, but that did not really help. 

Popular papers in the field of acoustics-based side-channel attacks usually use Mel Frequency Cepstrum Coefficients (MFCC)~\cite{mermelstein1976distance} for feature extraction, starting with one of the early attacks on a physical keyboard~\cite{zhuang2009keyboard}. 
However, MFCC was designed to replicate the human ear, and humans are not good at distinguishing tapping sounds on a screen. So MFCC is likely not optimal for tap classification. 
Quefrency (also known as Cepstrum), on the other hand, makes use of all the frequencies by calculating the Inverse Fast Fourier transform (IFFT) of their logarithm, but does not weight them according to the human ear's characteristics. 

So we use raw quefrency calculated on the first $128$ samples of audio data (after the tap) as an additional feature to our classifier. 
In fact, we combined quefrencies for the signals coming from one and more microphones. Throughout this paper, we will refer to three different feature 
sets: 1) Top -- refers to quefrency of a signal from the top microphone; 2) Botm -- refers to quefrency of a signal from the bottom microphone; 3) Top+Botm -- refers to Top features, concatenated to Botm features, concatenated to the estimated delay between the two signals and to quefrency of the difference between the two signals.   


For classification, we use Linear Discriminant Analysis (LDA), more specifically the one based on Singular Value Decomposition (SVD), as it showed the best results in our tests.

In the evaluation section (\fref{sec:evaluation}) we will present F1 scores for our attack. These are a better reference of test accuracy than a simple percentage of correctly guessed samples. F1 takes into the account both the precision (number of correct positive results divided by the number of all positive results) and the recall (number of correct positive results divided by the number of positive results). It is calculated as $$F1=2\times\dfrac{\text{precision}\times \text{recall}}{\text{precision}+\text{recall}}.$$

The macro F1 presented in this paper takes the average of the precision and recall reported for different labels.

%% file: sections/evaluation.tex
\section{Evaluation}\label{sec:evaluation}

\subsection{Experimental design}
\label{sec:experimentaldesign}
We developed an Android application where participants entered letters and words into an input field, or digits into a standard PIN pad.
In each case, the app displayed the text or digits to be entered by participants. The app worked in both landscape and portrait orientation; and had support for feedback mechanisms (sound, vibration). 
While the user performed the requested task, the application collected audio through the built-in microphones. 

As Android 5.1.1 prohibits the use of feedback mechanisms for PIN input, we tested them only for text input. 
We ran PIN-entry experiments only on the smartphone since the PIN pad has the same size on both tablet and smartphone. 
The PIN-code entry experiments were conducted only in portrait orientation because smartphones do not allow PIN input in landscape orientation.

We conducted the experiments outside of the lab environment in order to simulate real-life conditions more closely, thereby improving the validity of the study. Participants performed the task at the university in three different places: 
\first in a common room, where people chat and occasionally a 
coffee machine makes some Latte Macchiato along with loud sounds; \second in a reading room where people either type on a computer or 
speak in a semi-loud voice; and \third in the library where people are silent, but there are a lot of clicking sounds from nearby laptops. 
All three places had windows open, letting in ambient noise. 

As the speed of sound in air depends on its temperature, the data were only collected indoors during the day with air temperature ranging from 22--25 degrees Celsius and with no other user-started processes running in the background.
We used two LG Nexus 5 phones and a Nexus 9 tablet (all running  Android 5.1.1). 
The Nexus 5 measurements are 137.84mm $\times$ 69.17mm $\times$ 8.59mm and the Nexus 9 228.2mm $\times$ 153.7mm $\times$ 8mm. Both are fairly slim devices, with all of the internal components located relatively close to each other. Both devices have two microphones that can be accessed using the Android SDK. 
The main Nexus 5 microphone is located at the bottom of the device, the secondary one at the top. 
The Nexus 9 tablet is different, with one microphone 
at the bottom, and another one on the right side.


We used two LG Nexus 5 phones and a Nexus 9 tablet, which have a standard audio sampling rate. They both ran Android Lollipop which represented 35\% of devices at the time of the study and still represents 20\% in 2019. We suspect devices with higher sampling rates and better microphones may improve the attack, but we leave that for future work.

Due to the novelty of the attack itself, and the lack of a standard way of benchmarking side-channel attacks, the interpretation of our results is not straightforward. We want the reader to take away from the evaluation the following: 1) the attack on a small virtual keyboard performs about as well as to attacks on big physical keyboards; 2) the attack yields similar performance to previous virtual keyboard attacks based on other sensors such as the accelerometers and gyros, despite the microphone being much noisier.   

\subsection{Data Collection}
\label{sec:datacollection}

\begin{figure}
\centering
\begin{subfigure}{0.5\linewidth}
  \centering
  \includegraphics[height=200px]{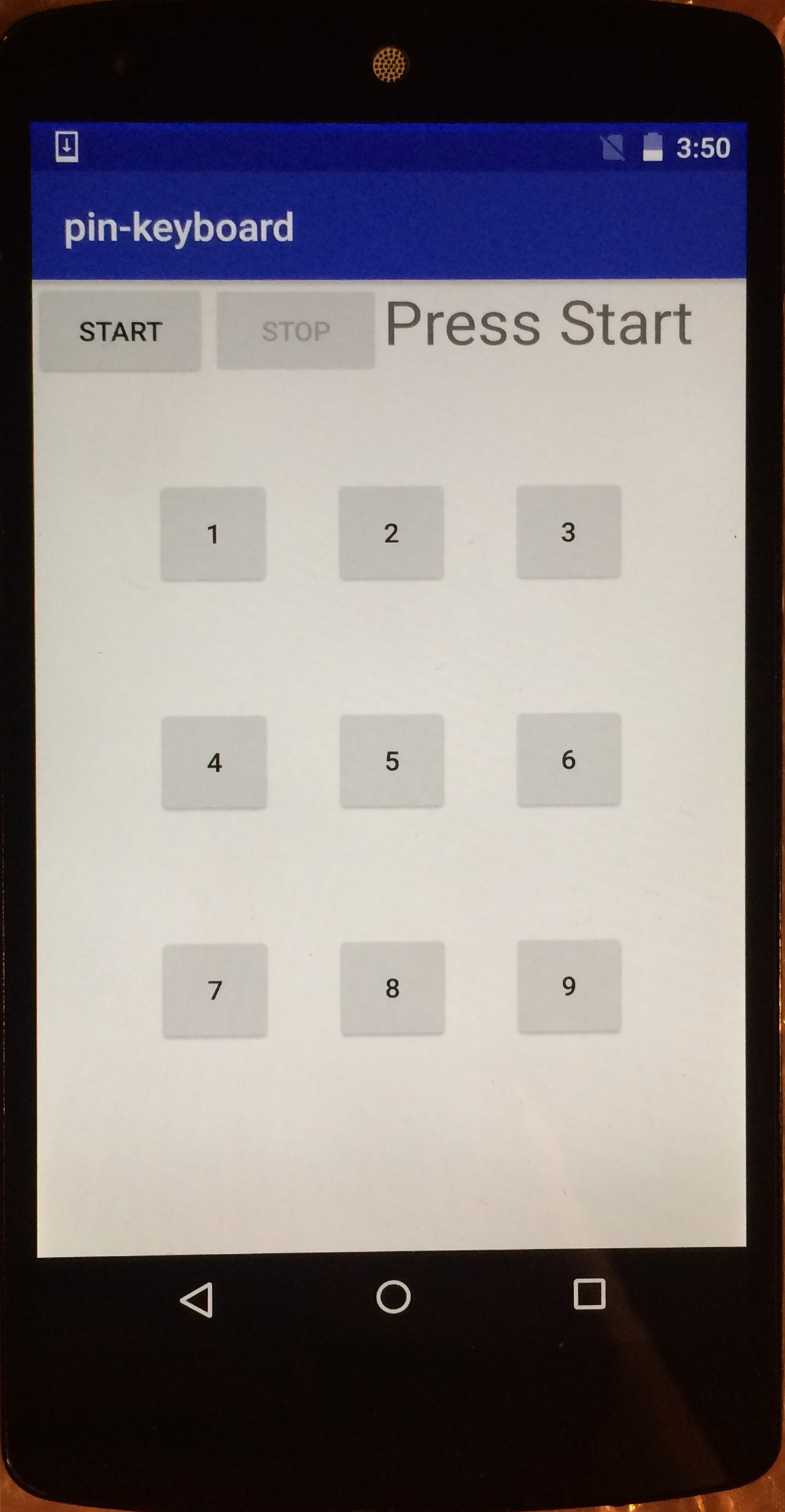}
  \caption{PIN-entry application}
  \label{fig:sub1}
\end{subfigure}%
\begin{subfigure}{.5\linewidth}
  \centering
  \includegraphics[height=200px]{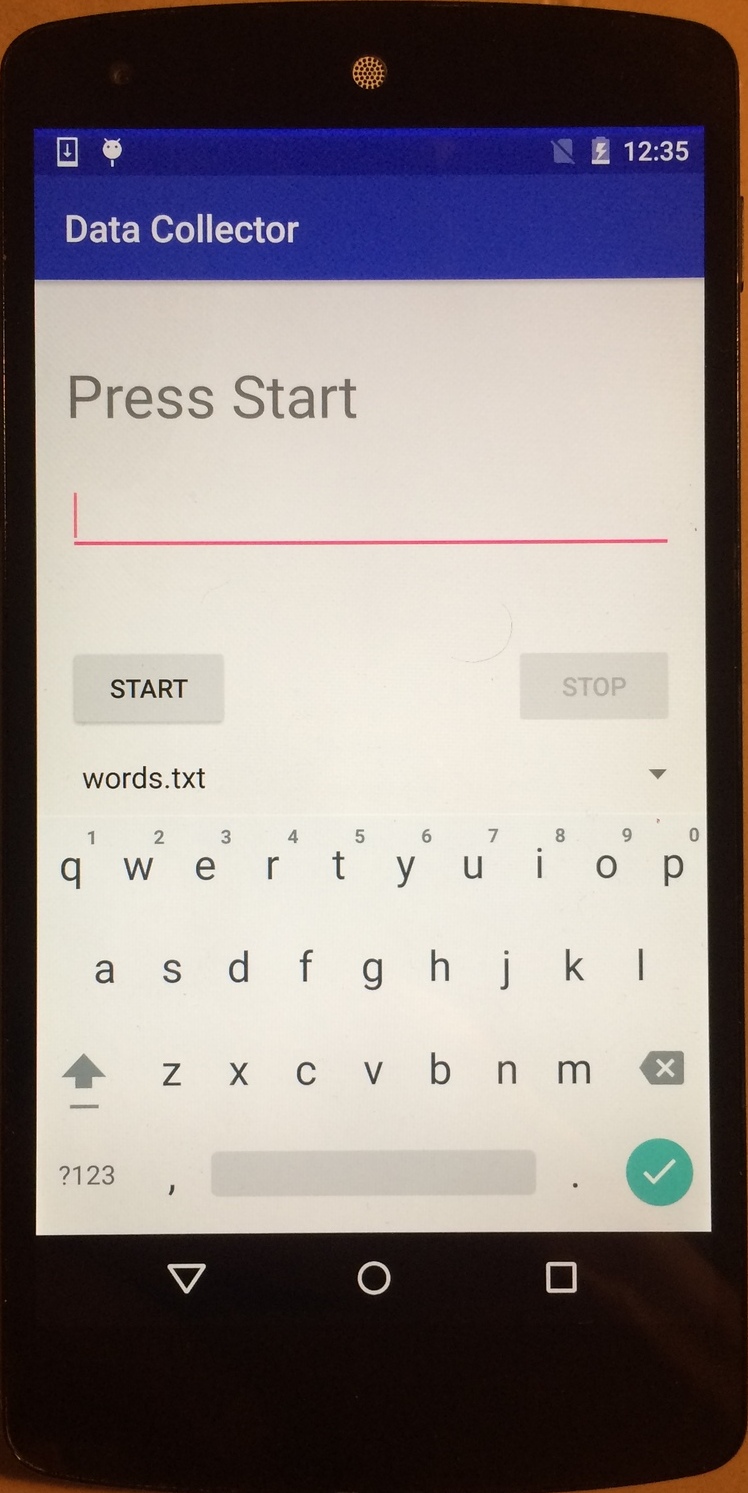}
  \caption{Text-entry application}
  \label{fig:sub2}
\end{subfigure}
\caption{Application developed for the experiments}
\label{fig:applications}
\end{figure}

We conducted four separate experiments. In the first, 10 participants were asked to press each of the nine digits (1 to 9) 10 times. The order of occurrence of the digits was randomised. In the second experiment, we asked 10 other participants to type 200 unique 4-digit PINs. In the third, we asked another group to type letters. The order of occurrence was also randomised. 
In the fourth experiment, participants were asked to type five-character words from the NPS chat corpus~\cite{Bird:2009:NLP:1717171}. 
A summary of all experiments for letter and word input is presented in~\fref{tab:keyboard-exper}. The applications developed for data collection are presented in \fref{fig:applications}.

\begin{table}[h]
\centering
\caption{Description of the experiments conducted.}
\label{tab:keyboard-exper}
\begin{tabularx}{0.7\linewidth}{lcccc}
\hline
what is being typed-in & orientation & feedback \\ \hline
\multicolumn{3}{l}{\textbf{Nexus 5 smartphone}} \\
200 most popular words (NPS) & portrait & no \\
200 most popular words (NPS) & landscape & no \\
100 most popular words (NPS) & landscape & sound \\
100 most popular words (NPS) & landscape & vibration \\
26 alphabet characters $\times$10 each & portrait & no \\
26 alphabet characters $\times$10 each & landscape & no \\
26 alphabet characters $\times$10 each & landscape & sound \\
26 alphabet characters $\times$10 each & landscape & vibration \\
9 digits $\times$10 each & portrait & no \\
4-digit PINS $\times$10 each & portrait & no \\
\multicolumn{3}{l}{\textbf{Nexus 9 tablet}} \\
100 most popular words (NPS) & portrait & no \\
100 most popular words (NPS) & landscape & no \\
26 alphabet characters $\times$10 each & portrait & no \\
26 alphabet characters $\times$10 each & landscape & no \\ \hline
\end{tabularx}
\end{table}

In total, we obtained data from 45 people, 21 female and 24 male. 
Two people refused to reveal their age; for the other 43 the mean age was 24.4 years and the range was between 21 and 32 years. Overall, we collected about 30 hours of audio. 


For each audio recording, we extracted the taps as explained in \fref{sec:background}. 
We then used two methods to test the accuracy of the attack. In the first, we used 70\% of the data for training the model (the ``train'' set), and the remaining 30\% for testing (the ``test'' set). Those datasets may contain data from the same users. 
In the second method, we performed the leave-one-subject-out (LOSO)  evaluation. Basically, the model is trained on N-1 participants and tested on 1; the results are averaged over all participants. 
In our threat model, we made an assumption of availability of records of 100 taps per character/digit prior to the attack. 
In the evaluation, this was implemented through random under-sampling prior to training of the training set -- we randomly dropped the samples from the original training dataset to contain 100 samples.
In the cross-validation (LOSO on top of k-fold), classifiers were run five times each and the results were averaged. Every k-fold run was performed on newly under-sampled data with new classifiers. Similarly, in the word prediction tasks, the training data was under-sampled every time. However, we also regenerated the word sequences every time from each of the participants of the experiments. 

\subsection{Ethical considerations}

This study went through the standard ethics review process in our institution and closely followed the departmental guidelines for data collection from human participants. The data were collected on a voluntary basis with no payment to participants. It was necessary to ensure that the participants were not harmed in any way and that no sensitive information was leaked. For these purposes, a number of measures were taken.

First, each of the volunteers was required to read and sign the consent form, which described the experiment. 
Consent was also obtained verbally; we explained the aim of the study to the participants to ensure they understood what type of data they were handing to us.

Second, no Personally Identifiable information (PII) were collected in our study, like name, date of birth or email address. The volunteers were only asked to type artificial data during the experiments which were randomly generated by us.

Third, we made sure that the data collected were impossible to trace back to each individual participant. 

Fourth, the data were stored on a secured computer, to which only the researchers in charge of analysis have access.

\subsection{Single digit inference}

\begin{table}[h]
\centering
\caption{Single digit classification accuracy on the Nexus 5 smartphone. Macro F1 is reported. Additional paper reported accuracies are presented.}
\label{tab:phone_pin_classification}
\begin{tabularx}{\linewidth}{@{}lllllll@{}}
\toprule
Feature-set          & train & test            & LOSO           \\ \midrule
Botm               & $34\pm4$ & $34\pm5$          & $31\pm22$          \\
Top              & $56\pm8$ & $55\pm5$          & $56\pm30$          \\
Top + Botm & $63\pm8$ & $\mathbf{64\pm6}$ & $\mathbf{66\pm37}$ \\ \midrule
Simon and Anderson~\cite{Simon:2013:PSI:2516760.2516770} & $ - $ & $26\%$ &-\\
Sun et al.~\cite{sun2016visible} & $ - $ & $38\%$ & -\\
Xu et al.~\cite{Xu:2012:TIU:2185448.2185465} & $ - $ & $40\%$ & - \\
Cai et al.~\cite{Cai:2012:PMB:2368367.2368385} & $ - $ & $50\%$ & -\\
Our best double & $ - $ & $\mathbf{65\%\pm3}$ & -\\
\bottomrule
\end{tabularx}
\end{table}

The train, test, and subject-independent (\ie LOSO) accuracies for single digit classification on the first attempt are presented in \fref{tab:phone_pin_classification}. 
With a single microphone, we correctly predict F1=$34\pm5$ and F1=$55\pm5$ of the PINs for the bottom (Botm) and the top microphones respectively -- 
compared to $1/9=11\%$ (although not a valid comparison between percentages and F1, it should still give an intuition about performance) for a random guess. Using two microphones, the accuracy improves to F1=$64\pm6$.

For the subject-independent case (LOSO), the standard deviation is higher. In the worst case, with the Botm microphone only, 
digits are predicted slightly better than a random guess; but in the best case almost half of the digits are correctly predicted on the first attempt. Results for the Top microphone are higher still, twice as good in the worst case and 85\% in the best case scenario. 
With two microphones, we correctly predict 3 times better than a random guess in the worst case; and 100\% 
of the digits in the best case.

\begin{figure}
    \centering
    \includegraphics[width=0.85\linewidth]{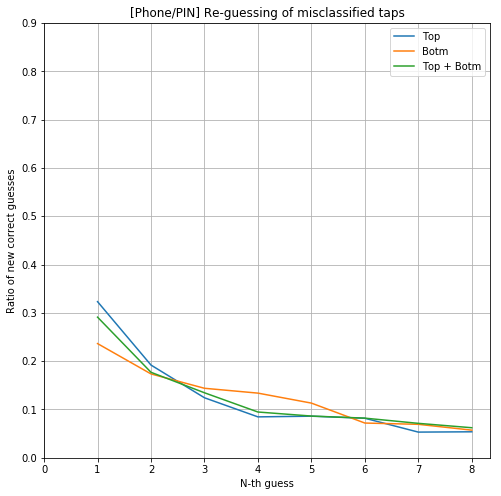} 
    \caption{The percentage of fixed errors with every new attempt.} 
    \label{fig:phone_pin_fixes}
\end{figure} 

When a digit is predicted incorrectly, the classifier is quick to rectify it (the guesses are ordered by the confidence value associated with a prediction yielded by the classifier). 
For example, with two microphones -- as shown in \fref{fig:phone_pin_fixes} -- 30\% of incorrect predictions are fixed after the first attempt with a further 20\% fixed at the second attempt. The rate of fixing then flattens out. 
This means that, in three attempts, the classifier correctly predicts more than 80\% of the digits. 

\begin{figure}
    \centering
    \includegraphics[width=0.85\linewidth]{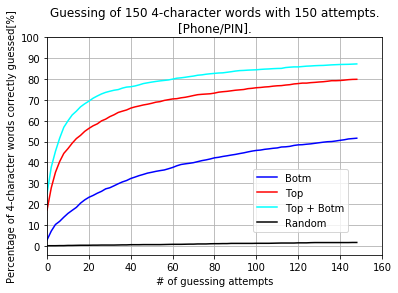} 
    \caption{Performance of the classifier trying to predict 200 random PIN-codes of size 4 collected from the Nexus 5 
    smartphone in landscape orientation based on the taps obtained from 20 people. The results are shown from 30 runs 
    with the average presented, each based on the mid-performing classifier.} 
    \label{fig:phone_pin}
\end{figure} 

\subsection{PIN inference}
The results for 4-digit PIN predictions are shown in \fref{fig:phone_pin}.
We consider 150 random PINs from 20 people as explained in \fref{sec:datacollection}. 
With a single microphone, we can recover 54\% of PINs after 10 attempts.
We compare our attack with previous work in \fref{tab:tdoa_nn}. It performs at least as well as some of them, and often better. For example, with two microphones, we recover 91/150 4-digit PINs in just 20 attempts -- without knowing anything about PIN distribution or the timing between subsequent taps~\cite{Bonneau12abirthday, pin_analysis_2012}. 

\begin{table}[h]
\centering
\caption{PIN Attack performance comparison. We report the best performing classifiers in single and double configurations. }
\label{tab:tdoa_nn}
\begin{tabularx}{\linewidth}{@{}llll@{}}
\toprule
Attack by & set size & $10^{th}$ try  & $20^{th}$ try \\ \midrule
Our best single   & 50 & 42\% & 50\% \\
Aviv et al.~\cite{Aviv:2012:PAS:2420950.2420957} & 50 & 55\% & - \\
Our best double & 50 & \textbf{55}\% & 61\% \\
Simon and Anderson~\cite{Simon:2013:PSI:2516760.2516770} & 50 &  61\% & 84\% \\
Spreitzer~\cite{Spreitzer:2014:PSE:2666620.2666622} & 50 & 79\% & - \\
Shukla~\cite{Shukla:2014:BYH:2660267.2660360} & 50 & 94\% & - \\
\midrule
Our best single   & 100 & 41\% & 49\% \\
Simon and Anderson~\cite{Simon:2013:PSI:2516760.2516770} & 100 &  48\% & 58\% \\
Our best double & 100 & \textbf{51}\% & 59\% \\
\midrule
Our best single   & 150 & 40\% & 48\% \\
Simon and Anderson~\cite{Simon:2013:PSI:2516760.2516770} & 150 &  44\% & 53\% \\
Our best double & 150 & \textbf{52}\% & 61\% \\
\midrule
Simon and Anderson~\cite{Simon:2013:PSI:2516760.2516770} & 200 &  40\% & 53\% \\
Our best single   & 200 & 43\% & 48\% \\
Our best double & 200 & \textbf{53}\% & 61\% \\
\bottomrule
\end{tabularx}
\end{table} 

\subsection{Letter \& Word inference}

In this section, we present the prediction results for the 26 letters of the English alphabet for both phone and tablet (\fref{tab:keyboard_res}). On the phone, we outperform a random guess by a factor of three if we access a single microphone for both LOSO and test. Audio from the top microphone yields better results than audio from the bottom microphone. We are not sure why this happens, but we suspect it may have to do with the way people hold the phone. Surprisingly, we found that portrait mode yields better results than landscape mode for the smartphone, while for the tablet, there is no noticeable difference between portrait and landscape modes. We believe this can be explained by the way we type on these devices: people usually put the tablet on their lap or a table in front of them, whereas they hold the phone in their hand and this may partially block the bottom microphone. When using two microphones, we obtain the best results, with an improvement of a factor of $1.5$ over a single microphone.

To compare our results to previous work, we followed the same evaluation methodology presented in \cite{liu2015good}, \cite{Berger:2006:DAU:1180405.1180436}, \cite{sun2016visible}, \cite{Marquardt:2011:IDV:2046707.2046771}: we randomly selected 27 words of length 7-13 from the corn-cob dataset and attempted to classify them. There are certain differences between our work and these papers though, for example, the training set size, the number of participants and the training procedures may vary slightly. Two of the papers evaluated attacks on physical keyboards which have higher accuracy in general. Although direct comparison is hard, this still provides a valuable baseline for our work. We present the comparison in \fref{tab:eval_text}.

Simon and Anderson's video side channel outperforms our attacks for a single microphone on the phone but shows similar results for the tablet. This suggests it is harder to perform acoustic attacks on a smartphone virtual keyboard as the distance between ``buttons'' is very short. For the tablet and when using two microphones, we obtain similar results as for attacks on physical keyboards.

\begin{table*}[!t]
\centering
\caption{Single character classification accuracy for 26 classes. Macro F1 value is presented.}
\label{tab:keyboard_res}
\begin{tabular}{*{6}{c|}c}
\hline
\multirow{1}{*}{Feature-set} & \multicolumn{3}{|c}{Phone} & \multicolumn{3}{|c}{Tablet} \\ \hline
& \multicolumn{1}{|c}{Train} & \multicolumn{1}{|c}{Test} & \multicolumn{1}{|c}{LOSO} & \multicolumn{1}{|c}{Train} & \multicolumn{1}{|c}{Test} & \multicolumn{1}{|c}{LOSO} \\ \hline
Portrait; Botm mic & $10\pm3$ & $11\pm1$          & $6\pm7$ & $26\pm3$ & $28\pm2$          & $22\pm16$  \\ \hline
Portrait; Top mic   & $16\pm3$ & $17\pm2$          & $10\pm9$  & $26\pm5$ & $25\pm2$          & $20\pm14$ \\ \hline
Portrait; both mics & $23\pm4$ & $\mathbf{23\pm3}$ & $\mathbf{12\pm15}$ & $44\pm3$ & $\mathbf{45\pm3}$ & $\mathbf{37\pm24}$ \\ \hline
Landscape; Botm mic & $10\pm3$ & $11\pm2$          & $7\pm8$ & $26\pm3$ & $26\pm2$          & $21\pm19$ \\ \hline
Landscape; Top mic   & $13\pm2$ & $16\pm3$          & $9\pm14$ & $24\pm3$ & $24\pm2$          & $19\pm12$ \\ \hline
Landscape; both mics & $20\pm4$ & $\mathbf{23\pm3}$ & $\mathbf{13\pm20}$ & $43\pm3$ & $\mathbf{44\pm1}$ & $\mathbf{37\pm27}$ \\ \hline
Random attack & $3\pm1$ & $3\pm1$ & $3\pm1$ & $3\pm1$ & $3\pm1$ & $3\pm1$ \\ \hline
\end{tabular} 
\end{table*}




\begin{table}[h]
\centering
\caption{27 corn-cob words of size 7-13 benchmark. We report the best performing classifiers in single and double configurations. }
\label{tab:eval_text}
\begin{tabularx}{0.6\linewidth}{@{}lll@{}}
\toprule
Attack by & 10-attempts & 50-attempts \\ \midrule
Phone/best single & 21\% & 30\% \\
Phone/best double & 25\% & 34\% \\
Marquardt et al.~\cite{Marquardt:2011:IDV:2046707.2046771} & 43\% & 56\% \\
Berger et al.~\cite{Berger:2006:DAU:1180405.1180436} & 43\% & 73\% \\
Tablet/best single & 43\% & 55\% \\
Liu et al.~\cite{liu2015good} & 63\% & 82\% \\
Sun et al.~\cite{sun2016visible} & 63\% &  93\% \\
Tablet/best double & 70\% & 80\% \\
\bottomrule
\end{tabularx}
\end{table}

\subsection{Word inference in real-world use}\label{sec:real}
The evaluation we presented in the previous section was motivated by previous papers.  However, in the real world it is much harder to actually perform attacks for a number of reasons, one of which is the lack of rich enough dictionaries and text corpora. 


This paper, however, suggests that one can perform inference attacks in such a situation. When mistakes are made, most of them get fixed quickly, suggesting that instead of dictionary search one can focus on a subset of predictions in which the classifier is the most confident, allowing some word to be guessed correctly.  

\begin{figure}
    \centering
    \includegraphics[width=0.85\linewidth]{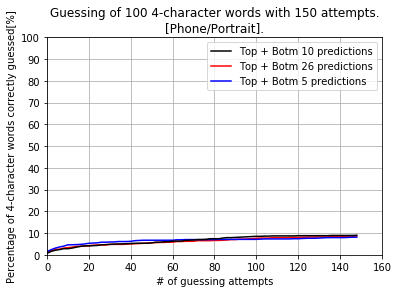} 
    \caption{Performance of the classifier trying to predict 100 most common words of size 4 collected from the Nexus 5 smartphone in portrait orientation with different combination depths based on the taps obtained from 20 people. The results are shown from 30 runs with the average presented, each based on the mid-performing classifier.} 
    \label{fig:phone_diff_comb}
\end{figure} 

Figure~\ref{fig:phone_diff_comb} shows the performance of word guessing based on a different prediction policy, with guesses going to depth 5 and 10. As can be seen, the overall accuracy stayed the same, despite a massively reduced prediction space. The space was reduced from $26^4$ to $10^4$ and $5^4$ or by 45 and 731 times respectively.  

Finally, in the real world most of the words in the dictionaries are extremely rare. To reflect this in our evaluation, we used the NPS chat corpus accessed through the NLTK framework~\cite{forsyth_lin_martell_2010, Bird:2009:NLP:1717171}, a real-world chat dataset. We then evaluated our attack on the most common 100 words of this dataset. To do that, we followed the approach of~\cite{Simon:2013:PSI:2516760.2516770} and synthesised words randomly from the collected single tap sounds. In addition to the classifier presented in \fref{sec:background}, we also trained an N-gram model using the Brown text corpus available through the NLTK framework~\cite{lindebjerg_1997, Bird:2009:NLP:1717171}. More precisely, to incorporate the information from the N-gram model into the predictions, we used the following function $$P_{combined}(pred) = P_{word}(pred)*P_{n-gram}(pred),$$ where $$P_{word}(pred) = \prod_{l=1}^{n} \mathit{p}(pred_{l})$$ and $$P_{n-gram}(pred) = \prod_{l=1}^{(n-k)} \mathit{p}(pred_{l:l+k}).$$ Here $pred_{l:l+k}$ is the subword from character on index $l$ to character on index $l+k$, and $p$ is the n-gram provided probability.

\begin{figure}
    \centering
    \includegraphics[width=0.85\linewidth]{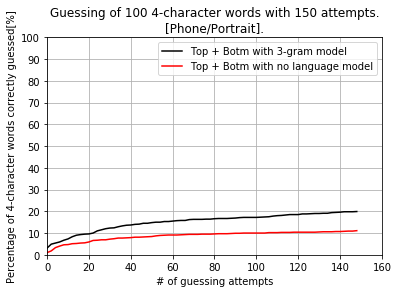} 
    \caption{Performance of the classifier trying to predict 100 most popular words of size 4 collected from the Nexus 5 smartphone in portrait orientation based on the taps obtained from 20 people. The results are shown from 30 runs with the average presented, each based on the mid-performing classifier.} 
    \label{fig:phone_word_real}
\end{figure} 

The results are presented in~\fref{fig:phone_word_real}. Without the language model, about 6\% of 4-letter words are correctly brute-forced in 20 attempts. The use of a 3-gram model~\cite{solomonoff1957inductive, 6773263} improves those results to 10\%. This is encouraging because it suggests that for real sentences, language models will help even more.

%% file: sections/countermeasures.tex
\section{Countermeasures} \label{sec:cm}
There are a number of ways to mitigate the acoustic attack presented in this paper. 
These can be implemented at different levels of the software and hardware stack.

At the hardware level, a radical solution is to remove the microphone from the handset or tablet. If projects like the Google Ara~\cite{atap_2014} gain traction, this may become viable for certain users. But this will never be acceptable for everyone, as we rely more and more on voice to interact with smart devices.

One could introduce physical switches to phones, so users could turn the microphone off. 
But this would introduce usability issues for most of us: imagine having to physically enable the microphone every time you answer a voice call! Press-to-talk radios were progressively replaced by voice-operated microphones since the technology became available fifty years ago, and the trend in phones has been to have fewer buttons over time. 

Many other possibilities go too strongly against the grain of commercial competition. One could lower the microphone sampling frequency, but as vendors compete for the best video recording, this is unlikely to be acceptable. An additional glass layer on top of the screen could absorb most of the finger impact, but would not appeal to vendors fighting to make devices slimmer and lighter. 

The most usable approach at the level of hardware design might be to make live microphones obvious, perhaps by having a diode connected to the energy rails of the microphone itself, lighting up whenever it's in use, and forming an icon that represents a microphone. The obvious objection is that a live-mike icon would often be ignored, like most security warnings. However, it would give some value to some users, and although it would carry a small cost in hardware, something similar could be implemented in the OS for zero marginal cost.


At the operating-system level, one might also try to prohibit audio recording during data entry. But this would break many apps, for example when users type during a voice call. 

A more sophisticated approach would be to offer a properly-engineered PIN entry facility, which when called by one application would temporarily blank, and/or introduce noise into, the microphone channel seen by other apps. This approach should logically be extended to other sensors that can be used to harvest PINs via side-channels such as the accelerometer, gyro and camera. If restricted explicitly to PIN entry, it might be used sufficiently sparingly that its interference with other apps might well be tolerable.

As the next step, a "secure text entry" facility could be provided for apps to solicit passwords and other authentication information. But thought would need to be given to possible interactions, for example between banking apps that insist on secure password entry and navigation apps that insist on continuous access to accelerometers and gyros. If secure text entry came to be the default for messaging apps that offer end-to-end encryption, the scale of use might start to cause problems elsewhere in the app ecosystem.

More subtle degradation should be explored, such as for the OS to introduce timing jitter, or decoy tap sounds, into the microphone data stream. This way an attacker could not reliably identify tap locations. As the taps themselves are pretty unnoticeable for humans, this should not disturb disturb applications that run in the background and collect audio with user consent. 



At the application level, an app might itself introduce false tap sounds into the device, in order to jam and confuse any hostile apps that happen to be listening. Such tactical jamming is a low-cost countermeasure and as far as we can see may be the most feasible mitigation for a developer of (say) a banking app to deploy today. 




Overall, we favour two approaches: 1) reporting to the user which sensors are on; 2) introducing a secure attention sequence -- a new high-priority mode in the phone interaction pipeline for passwords or other sensitive text entry, in which all sensors except the touch screen are blocked. The first is fairly straightforward but raises issues of usability, liability and compliance, while the second requires more engineering but might provide higher assurance. Until these (or other mitigations) are implemented in the platform, app developers should consider the use of tactical jamming if PIN theft via side channels is ever deployed at scale.

%% file: sections/limitations.tex
\section{Limitations}

In our threat model, we used only the microphones to classify and detect finger taps, and we limited ourselves to 26 English characters and 9 digits. Once we move from password capture to the inference of more general text input, there are other characters to consider, such as the backspace, the space buttons, \etc These complicate the machine-learning models and will lead to lower prediction accuracy. But attackers could use other sensors too, such as the accelerometers and gyroscopes. So it would be of interest to evaluate the attacks that are possible using multi-sensor side channels, and measure the performance of different classifiers with different alphabets and in different SNR settings.

In our attack, we managed to recover letters in the subject-independent model). It is unclear why the attack generalises well; presumably people's fingers are more similar than their voices are. More research on this might be useful. Future research could also investigate the transferability of the attack to different types of device. However, the trend towards more and better-quality microphones suggests that attacks will only get better. 

%% file: sections/related_work.tex
\section{Related Work}

This paper fits into the area of side-channel attacks on smartphones. In particular, we present a novel acoustic side-channel. The microphone is not typically considered to leak information about the locations of taps on the screen, and acoustic attacks are relatively unexplored~\cite{Cai:2009:DAS:1592606.1592614}. 

Attacks on physical keyboards are not new. Asonov and Agrawal were among the first to demonstrate acoustic attacks on mechanical keyboards~\cite{asonov2004keyboard}. Their work was further improved by Zhuang et al.~\cite{zhuang2009keyboard} who showed that an attacker can deduce text even in unsupervised operation, and by Compagno et al. who showed how to infer text based on keyboard acoustics acquired through Skype~\cite{DBLP:journals/corr/CompagnoCLT16}. 

But what about virtual keyboards on the screens of smartphones or tablets? Cai and Chen showed that an attacker can use motion sensors to recover PIN codes entered by a user with a 70\% recovery rate~\cite{Cai:2012:PMB:2368367.2368385}. Their threat model is very similar to ours -- they assume a malicious app residing on the victim's phone with sensor access. 

So is it enough for a security-conscious user to stop apps using the motion sensors? Simon and Anderson reported an attack using camera movement to measure handset orientation, instead of the accelerometers and gyros, and using the microphone for the tap detection~\cite{Simon:2013:PSI:2516760.2516770}. This let them recover PINs with reasonable probability. 

Our attack continues this line of research. It's not enough to deny apps permission to use motion sensors and the camera; to stop PIN leakage completely you have to deny access to the microphones too. That will be infeasible for most users and many apps.

Our attack exploits the double microphones found in many modern smartphones. These have been used previously to perform acoustic attacks, but not on the smartphone itself. For example, Liu et al. presented a system for keyboard press inference using the multiple microphones of a nearby smartphone~\cite{Liu:2015:SKM:2789168.2790122}. Similarly, Zhu et al.  present several context-free attacks on keyboard emanations using only time difference of arrival information~\cite{Zhu:2014:CAU:2660267.2660296}, and show that multiple smartphones can increase the accuracy of the results they obtain.

%% file: sections/conclusion.tex
\section{Conclusion}\label{sum}

We present a new acoustic side-channel attack on touch-screen devices such as smartphones and tablet computers. We use the device's own microphones to infer text entry from the sounds made by finger taps on the screen.

We showed that the attack can successfully recover PIN codes, individual letters and whole words. On a Nexus 5 smartphone in portrait orientation, the attack was able to recover most PINs and some words; specifically, 146 of 200 4-digit PINs after 10 attempts, and 8 out of 27 random words from the corn-cob dataset of size 7-13 after 20 attempts. 

We also showed that it is possible to achieve high accuracy in a subject-independent setting, so one can attack users' smartphones remotely with a classifier pre-trained on other peoples' devices. 

Our attack achieved an even better accuracy on tablets than on smartphones. For example, on a tablet, we recovered not just some words but most words; specifically, 19 words out of 27 after 10 attempts. 

To assess the practical threat, when guessing words from a set of words commonly used in text messaging, we were able to recover 80 words out of 200 after 20 attempts. Furthermore, insights into the way mistakes are made allowed us to develop a model for an effective attack that would use a language model on both letter and word levels. 


In conclusion, we have shown a new acoustic side-channel attack on smartphones and tablets, and described how to exploit it effectively.

Finally, we have discussed what mitigations are possible, both for application writers now, and for future versions of mobile phone operating systems. 